\documentstyle[12pt]{article}
\begin{document}
\begin{center}
\underline{QUANTUM LATTICE FLUCTUATIONS AND LUMINESCENCE}\\
\underline{IN $C_{60}$}\\
\ \\
Barry Friedman\\
Department of Physics and Texas Center for Superconductivity\\
University of Houston\\
Houston, Texas 77204-5504\\
and \\
Department of Physics\\
Sam Houston State University\\
Huntsville, Texas 77204-5504\footnote{Present and permanent address.}\\
\ \\
Kikuo Harigaya\\
Fundamental Physics Section, Physical Science Division\\
Electrotechnical Laboratory\\
Umezomo 1-1-4, Tsukuba, Ibaraki 305, Japan\\
\end{center}
ABSTRACT\\
\indent
We consider luminescence in photo-excited neutral $C_{60}$ using the
Su-Schrieffer-Heeger model applied to a single $C_{60}$ molecule.
To calculate the luminescence we use a collective coordinate method
where our collective coordinate resembles the displacement of the carbon
atoms of the Hg(8) phonon mode and extrapolates between the ground state
"dimerisation" and the exciton polaron.  There is good agreement for the
existing luminescence peak spacing and fair agreement for the relative
intensity.  We predict the existence of further peaks not yet resolved
in experiment. 
\newpage
\indent
        In a recent experiment, Matus, Kuzmany and Sohmen\cite{matus} measured
luminescence from $C_{60}$ films and interpreted their results in terms of an
exciton polaron.  The main purpose of this letter is to support the above
author's interpretation with some simple calculations and to clarify a few
points.

        We interpret the luminescence in $C_{60}$ within the Su-Schrieffer-
Heeger (SSH)\cite{su}
 model applied to a single $C_{60}$ molecule\cite{har}. That is, we work with
the Hamiltonian
\begin{equation}
H=\sum_{<i,j>} \lbrace \,-(t-\alpha(X_{ij}-a)) \sum_{\sigma} [
c_{i,\sigma}^{\dag} c_{j,\sigma} + H.c.] + \frac{K}{2}(X_{ij}-a)^{2}
 \rbrace +
U\sum_{i} n_{i\uparrow}n_{i\downarrow} +1/2\sum_{j}m(d\vec{r_{j}}/dt)^{2} .
\end{equation}
Here  $\vec{r_{j}}$ is the
cartesian coordinate of the jth carbon atom, $a$ is the
bare carbon-carbon bond length and $X_{ij}$ is the distance between the ith
and jth carbon atoms.  Our experience with luminescence in conducting
polymers\cite{barry}
 leads us to two approximations: 1) We neglect intermolecular
hopping (we discuss this assumption later). 2) We set U=0 (explicit
electron-electron interaction is neglected).  When considering vibrational
properties of conducting polymers this is a good zeroth order approximation.

        A number of authors\cite{har}
 have treated the above Hamiltonian using these
approximations under the further restriction that the lattice (the 60
carbon atoms) is treated classically.  For our purposes, the most important
result of these studies is the formation of the exciton polaron when an
electron is promoted from the $H_{u}$ orbital to the $T_{1u}$ orbital.  That
is,
the lattice distorts in the sense that the dimerisation (
the difference in the two different bond
lengths) is virtually destroyed on a ring circling the $C_{60}$ molecule.
Concurrent with the lattice distortion, two electronic states are pulled
into the gap.  Electrons occupying these states live predominantly on the
distorted part of the lattice (i.e. the ring).  It is important to note
that the exciton polaron does not break the inversion symmetry of the
system.

    Our picture of luminescence in $C_{60}$ is then as follows: By a complex
dynamical process the photoexcited
$C_{60}$ molecule evolves into the state where one
electron occupies the lower gap energy level and one electron occupies the
upper gap energy level.  By spontaneous emission the system then decays into
the
electronic ground state and a possibly excited lattice vibrational state.  A
complication here is that the exciton polaron does not break parity and the
transition from the highest occupied molecular orbital to the lowest
unoccupied orbital is dipole forbidden.

        The above picture entails that we must treat the lattice quantum
mechanically.  We continue to work within the adiabatic approximation. However
it is difficult
to work in the adiabatic approximation without further approximation
 since the lattice has 180 degrees of freedom.  We therefore adopt
the collective coordinate method\cite{barry}
, reducing the problem of 180 degrees of
freedom to a single judiciously chosen collective coordinate.  This method has
been used with reasonable success to calculate the absorbance and luminescence
in nondegenerate conducting polymers\cite{barry}.

        The key ingredient for the collective coordinate method is, not
surprisingly, a good choice for the collective coordinate.  In conducting
polymers, a good collective coordinate has been found to be a one parameter
family of lattice configurations that extrapolates between the ground state and
first excited state classical lattice configurations.  In the case of $C_{60}$,
the ground state of the lattice is dimerised, the bonds separating hexagons
from
hexagons, h-h bonds, have a length $l_{1}$ and the bonds separating pentagons
from hexagons, p-h bonds, have a length $l_{2}$.
  Experiment gives a value of $l_{1}$ = 1.40 ${\rm\AA}$
and $l_{2}$=1.45${\rm\AA}$.  For the first excited state calculations with the
 SSH  model
tell us the bond length pattern is largely the same as that of the ground
state other then on a ring circling the molecule.  On this ring, consisting of
twenty carbon atoms, the difference in bond lengths between the h-h and h-p
bonds is suppressed.  We therefore choose a collective coordinate $u$ so that
carbon atoms not on the ring are fixed and for $\vec{r_{i}}$ on the ring
\begin{equation}
\vec{r_{i}}=(i\tilde{a} + (-1)^{i}u) \hat{x},
\end{equation}
where the index $i$ ($1 \leq i \leq 20$) labels the carbon atoms on the ring.
Here we treat the ring as a chain along the x-axis with periodic boundary
conditions.  This is a simple and reasonable approximation since the quantity
entering into the total energy is the bond length difference.  The parameter
$\tilde{a}$, the renormalised bond length is equal to $a$ - average bond length
shrinkage, where $a$ is the unrenormalised bond length. (For the parameter
values we have adopted the h-h shrinkage $\approx$ .11 ${\rm\AA}$
  and the p-h shrinkage $\approx$ .16 ${\rm\AA}$,
the average shrinkage being therefore $\approx$ .14 ${\rm\AA}$).
 In this choice of collective
coordinate, $u$=0 approximates the distortion of the exciton polaron and $u$=
(p-h bond shrinkage - h-h bond shrinkage)/4 approximates the ground state
dimerisation.  We emphasize that our collective coordinate is not an
unreasonable approximation to the Hg(8) phonons\cite{iwasa}
 obtained from microscopic
calculations\cite{sankey}.  In particular, these calculations show that the
twenty carbon atoms on the ring are displaced almost parallel to the x-axis.

        We proceed to examine the consequences of our collective coordinate.
The lattice kinetic energy (eq. 1) then reads in terms of the collective
coordinate $u$
\begin{equation}
   1/2\sum_{i=1}^{20} m(d\vec{r_{i}}/dt)^{2} =  1/2\sum_{i=1}^{20}
m(du/dt)^{2}=
1/2 M(du/dt)^{2} .
\end{equation}
               where $M$=20$m$ is the mass of 20 carbon atoms.  With
this kinetic energy it is easy to write down the collective coordinate
Schr\"{o}dinger Equation
\begin{equation}
-\frac{\hbar^{2}}{2M}d^{2}\psi/d^{2}u + V(u)\psi = E\psi
\end{equation}
V($u$), the adiabatic potential energy
 is computed as the total energy of the SSH
Hamiltonian for a fixed value of u. In figure 1 we have plotted the adiabatic
potential energy for the electronic ground state and first excited state. Of
course, to obtain such a curve we have used parameter values for t,$\alpha$
  and K. For
polyacetylene, we find
  t=1.35 eV, $\alpha$=7.0 eV/${\rm\AA}$ and K=53 eV/${\rm\AA^2}$
work rather well to
reproduce the experimental optical properties and dimerisation.  If we use
these parameter values for $C_{60}$ we obtain a slightly too large bond length
difference (ie. .06 ${\rm\AA}$
) and a somewhat too large optical gap ($\approx$2.2 eV).
We have consequently adjusted
$\alpha$ to give the proper dimerisation by decreasing
$\alpha$ to 6.3 eV/${\rm\AA}$.  Such a value of $\alpha$
 reduces the naive gap (see below)
to 1.96 eV in better agreement with the experimental optical gap of 1.9eV
\cite{matus}.
        By solving the collective coordinate Schr\"{o}dinger
 equation in the lower
adiabatic potential we obtain a series of discrete levels separated by about
.21 eV.  The energy differences between these levels should correspond to the
energy difference between vibronic peaks in luminescence. There is consequently
reasonable agreement between our calculation and experiment since Matus,
Kuzmany and Sohmen\cite{matus}
 report prominent peaks at 1.70 and 1.52 eV, that is an energy
difference of .18 eV. According to our calculation, there should be
additional equally spaced peaks.  We attribute the absence of such peaks (which
we calculate to have smaller intensity, see below) to experimental
uncertainties, material problems etc..  An earlier measurement of luminescence
in $C_{60}$
films by Reber et al.\cite{reber}
 , in fact, seems to resolve 3 peaks differing in
energy by $\approx$.17 eV. Preliminary results of Iwasa et al.\cite{iwasa} also
indicate the presence of additional peaks.
Our value of energy level differences of .21 seems
relatively insensitive to parameter choices, for example if we let
$\alpha$=7.0 eV/${\rm\AA}$
we get a level spacing of .2 eV.

        A consequence of our theory is that the energy difference between peaks
in luminescence should depend on the isotope of carbon present in the $C_{60}$
molecule.  Since the adiabatic potential is very close to harmonic (for the
lower curve) the energy spacing depends on the mass of the carbon atom
 $m$ like
$m^{-1/2}$.
 It therefore may be interesting to do experiments on luminescence in
$C_{60}$ films made using $C^{13}$\cite{lieber}.

        We next turn to the intensity of the luminescence.  The intensity of
the luminescence is proportional to (with an energy independent constant)
\begin{equation}
\omega^{4}\mid\int_{-\infty}^{\infty}\psi^{*}_{i}(s)\psi_{f}(s)Q(s)ds\mid^{2}.
\end{equation}
where $\psi_{i}$, $\psi_{f}$
are the initial and final vibrational wavefunctions and Q(s) is the
electronic matrix element for the lattice configuration with fixed collective
coordinate s.  The difficulty here is that in the dipole approximation for an
isolated $C_{60}$ molecule
 (and our collective coordinate) Q(s) is zero.  We expect
that oxygen-impurities,coexisting $C_{70}$ and/or solid state physics effects
 (other $C_{60}$ molecules) will make
Q(s) nonzero even in the dipole approximation.  Such effects are not simple to
estimate, fortunately Q(s) is probably only weakly s dependent.  Therefore, we
can treat Q(s) as a constant and pull it out of the integral.  In calculating
relative intensities Q then doesn't enter, we need only consider the quantity
\begin{equation}
\omega^{4}\mid\int_{-\infty}^{\infty}\psi^{*}_{i}(s)\psi_{f}(s)ds\mid^{2}.
\end{equation}
A straightforward calculation yields figure 2.  In figure 2, the solid circles
are our calculation, while the solid curve is the experiment of Matus, Kuzmany
and Sohmen\cite{matus}  and the dashed curve is the experiment of Reger et al.
\cite{reber}.  Our calculation
is by no means in perfect agreement with experiment, it does however seem to be
not unreasonable especially for such a simple theory (Actually our calculated
relative intensities agree remarkably
 well with the experiment of Reger et al).

        Lastly, we consider absorption.  Our model for absorption is that the
lowest vibrational wavefunction in the first adiabatic potential makes a
transition to various vibrational wavefunctions in the second adiabatic
potential.  We have plotted the absorption in figure 3.  The most intense
absorption occurs at 1.9 eV.  This is in apparent agreement with the onset of
absorption reported in ref.[1].

        We have considered luminescence within a simple model.  Our model
agrees well with the existing experimental energy differences and there is fair
agreement with the relative intensity.  We predict that more extensive
experiments will see more peaks in luminescence and an isotopic shift for the
luminescence peaks if $C^{12}$ is replaced by $C^{13}$.

        This work was partially supported by the Sam Houston State University
research enhancement fund, the Robert A. Welch Foundation and by the Texas
Center for superconductivity at the University of Houston under Prime Grant No.
MDA 972-88-G-002 to the University of Houston from the Defense Advanced
Research Projects Agency and the State of Texas.  One of the authors (K.H.)
thanks Dr. S. Abe for many fruitful discussions.  We also acknowledge helpful
conversations with Professor W. P. Su (with K. H. at the Electrotechnical
 Technical Lab and at ICSM'92) and Dr. Y. Iwasa.

\newpage
\noindent
FIGURE CAPTIONS\\

\noindent
Figure 1. Adiabatic potential energy vs collective coordinate $u$.  For our
parameter values $u$=.0125${\rm\AA}$ corresponds to a dimerised
lattice and $u$=0 corresponds
to the exciton polaron.\\

\noindent
Figure 2. Relative intensity of luminescence vs energy. The large dots are our
calculation, the dashed curve is an envelop of the experimental data from Reber
et al.\cite{reber} and the solid curve
is from the experiment of Matus, Kuzmany, and
Sohmen\cite{matus}.\\

\noindent
Figure 3. Relative intensity of absorption vs energy.  The large dots are our
calculation.\\

\noindent
Note:  Figures will be sent by the conventional mail.  Please send your
request to harigaya@etl.go.jp.

\end{document}